\documentclass{sigchi}

\usepackage{balance}       % to better equalize the last page
\usepackage{graphics}      % for EPS, load graphicx instead 
\usepackage{caption}
\usepackage{subfig}
\usepackage[T1]{fontenc}   % for umlauts and other diaeresis
\usepackage{txfonts}
\usepackage{mathptmx}
\usepackage[pdflang={en-US},pdftex]{hyperref}
\usepackage{color}
\usepackage{booktabs}
\usepackage{textcomp}
\usepackage{stfloats}
\usepackage[none]{hyphenat}

\usepackage{microtype}        % Improved Tracking and Kerning
\usepackage{ccicons}          % Cite your images correctly!
\usepackage{todonotes}

\def\plaintitle{Modeling Human Visual Search Performance on Realistic Webpages Using Analytical and Deep Learning Methods}

\def\emptyauthor{}
\def\plainkeywords{Performance modeling; deep learning; scannability; convolutional neural network; webpage; visual attention.}

\makeatletter
\def\url@leostyle{%
  \@ifundefined{selectfont}{
    \def\UrlFont{\sf}
  }{
    \def\UrlFont{\small\bf\ttfamily}
  }}
\makeatother
\def\pprw{8.5in}
\def\pprh{11in}

\definecolor{linkColor}{RGB}{6,125,233}
\hypersetup{%
  pdftitle={\plaintitle},
  pdfauthor={\emptyauthor},
  pdfkeywords={\plainkeywords},
  pdfdisplaydoctitle=true, % For Accessibility
  bookmarksnumbered,
  pdfstartview={FitH},
  colorlinks,
  citecolor=black,
  filecolor=black,
  linkcolor=black,
  urlcolor=linkColor,
  breaklinks=true,
  hypertexnames=false
}

\toappear{\scriptsize Permission to make digital or hard copies of part or all of this work for personal or classroom use is granted without fee provided that copies are not made or distributed for profit or commercial advantage and that copies bear this notice and the full citation on the first page. Copyrights for third-party components of this work must be honored. For all other uses, contact the owner/author(s). \\
{\emph{CHI '20, April 25--30, 2020, Honolulu, HI, USA.} } \\
Copyright is held by the author/owner(s). \\
ACM ISBN 978-1-4503-6708-0/20/04. \\
http://dx.doi.org/10.1145/3313831.3376870}
\begin{document}

\title{\plaintitle}

\numberofauthors{3}
% \author{%
%   \alignauthor{Leave Authors Anonymous\\
%     \affaddr{for Submission}\\
%     \affaddr{City, Country}\\
%     \email{e-mail address}}\\
%   \alignauthor{Leave Authors Anonymous\\
%     \affaddr{for Submission}\\
%     \affaddr{City, Country}\\
%     \email{e-mail address}}\\
% }
\author{%
  \alignauthor{Arianna Yuan\footnotemark \\
    \affaddr{Stanford University}\\
    \affaddr{Stanford, CA, USA}\\
    \email{xfyuan@stanford.edu}}\\
  \alignauthor{Yang Li\\
    \affaddr{Google Research}\\
    \affaddr{Mountain View, CA, USA}\\
    \email{liyang@google.com}}\\
}

\maketitle
\footnotetext{\textsuperscript{*}This work was completed while the author was an intern at Google.}

\begin{abstract}
  Modeling visual search not only offers an opportunity to predict the usability of an interface before actually testing it on real users, but also advances scientific understanding about human behavior. In this work, we first conduct a set of analyses on a large-scale dataset of visual search tasks on realistic webpages. We then present a deep neural network that learns to predict the scannability of webpage content, i.e., how easy it is for a user to find a specific target. Our model leverages both heuristic-based features such as target size and unstructured features such as raw image pixels. This approach allows us to model complex interactions that might be involved in a realistic visual search task, which can not be easily achieved by traditional analytical models. We analyze the model behavior to offer our insights into how the salience map learned by the model aligns with human intuition and how the learned semantic representation of each target type relates to its visual search performance. 
\end{abstract}

% ACM Classfication



% Author Keywords

% Print the classficiation codes
% Please use the 2012 Classifiers and see this link to embed them in the text: \url{https://dl.acm.org/ccs/ccs_flat.cfm}

\section{Introduction}

Modeling human visual attention in interaction tasks has been a long-standing challenge in the field of human computer interaction \cite{borji2019saliency, tehranchi2018modeling, van2016towards, wu2018analysis, Walter:2015:AVA:2750858.2804255}. Building reliable models that can accurately estimate the difficulty of various visual search tasks has significant importance to user interface (UI) design and development. Traditional approaches for examining visual search involve usability tests with real human users that are time-consuming and expensive. A generalizable, predictive model can save time and cost for UI practitioners such as UI designers or UX researchers by offering them insights into visual search difficulty before testing with real users.

Many classic results on visual search have been established in the field. For instance, researchers have differentiated two types of visual search: feature search and conjunction search. Feature search is a visual search process in which participants look for a given target surrounded by distractors that differ from the target by a unique visual feature, such as orientation, color and shape, e.g., searching a triangle amongst squares. On the other hand, in conjunction search, participants look for a previously given target amongst distractors that share one or more visual features with the target \cite{shen2003guidance}, such as searching a red triangle amongst red squares and yellow triangles. Previous studies have shown that the efficiency (reaction time and accuracy) of feature search does not depend on the number of distractors \cite{mcelree1999temporal}, whereas the efficiency of conjunction search is dependent on the number of distractors present---as the number of distractors increases, the reaction time increases and the accuracy decreases \cite{treisman1980feature}. In addition, using well-controlled visual stimuli, previous studies showed that prior knowledge of the target greatly influence visual research time \cite{zhaoping2011clash}, indicating the interaction between top-down and bottom-up processing in visual search process.

Despite the robustness and simplicity of these early findings, they are not very practical and could only model search time in restricted laboratory settings. For example, the effect of number of distractors on visual search time in conjunction search is observed with oversimplified stimuli, e.g., geometric shapes or Gabor patches. This is very unrealistic in everyday visual search tasks, e.g., finding a booking button on a busy hotel homepage. Later modeling studies have attempted to simulate visual search in more realistic context, such as searching a target in webpages, menus and other graphical interfaces \cite{jokinen2020adaptive, fu2007snif,teo2012cogtool}. However, those studies usually require the extraction of a set of predefined visual features of the target and the candidate visual stimuli. Although those predefined, handcrafted features are indicative for search time, there is often useful information in the visual scene that cannot be captured by those handcrafted features and is missing in previous computational models.

On the other hand, recent advance in deep learning has demonstrated its great power in transforming many fields, such as speech recognition, machine translation and object detection \cite{johnson2017google, devlin2018bert, lecun2015deep}. One significant benefit of using deep learning is that a model will pick up on features that may not be evident yet important for prediction. Deep learning also has the capacity to capture the complicated interactions between different factors. Using deep learning for human performance modeling as well as solving other interaction problems has started to gain traction with the HCI community \cite{Pfeuffer:2018:AMG:3173574.3173862}. In this work, we present a method that takes advantage of both traditional modeling approaches and the popular deep learning method. Particularly, we combine existing heuristic-based, structured features that have long been used to model visual search tasks, together with the unstructured image features---raw pixels of a webpage---to predict human visual search time. 

To process a vast amount of information such as raw pixels of a webpage, we use convolutional neural networks (CNN) as our image feature extractors. CNN has been widely used in object recognition, object detection and visual question answering \cite{krizhevsky2012imagenet, ren2015faster, chen2015abc}. This powerful architecture allows us to capture the visual details on the webpage and the interaction between the webpage and the target without deliberately specifying any features. To make our experiments as realistic as possible, we focus on a common interaction task, where users search for a target item from a webpage, e.g., the ``Home'' button, the ``Read More'' tab, an image of a product, or a link to another webpage. The task involves diverse webpages that were collected from the Internet. The targets are from a rich set of categories, including images, text, links, buttons and input fields, which makes our search tasks rich and realistic but at the same time challenging to model. 

Because we want to provide a unified model to process the unstructured image data regardless of target type and webpage type, we build a neural network that only takes the raw pixels of the webpage and the target as inputs. Our neural network computes an attention map representing the places that the model should attend to given the target and the webpage. This attention map is then combined with various structured features, e.g., target position and target size, to predict the search time. The entire model
is trained end-to-end using back-propagation. The model outperforms baseline models that use only structured features. Importantly, it provides a way to integrate both structured and unstructured features and thus can benefit from both the traditional approaches and the deep learning methods.

The contributions of the paper are the following: 
\begin{itemize}
	\item We present a deep learning approach that is capable of predicting human visual search time on large-scale realistic webpages. It opens a new avenue for building predictive models that address a broader set of conditions, which complements previous work. 
	
	\item It provides a way to flexibly integrate both structured and unstructured features, which gives it unique strengths to not only fit the training data but also generalize to unseen data. It outperforms traditional models that use only structured features.
	
	\item We offer qualitative understandings about our model behavior by examining the attention map as well as other latent representations learned by the model. We found spontaneous emergence of sensible embeddings for target types, as well as attention map that matches human intuition. This enhances the interpretability of the deep learning model.
\end{itemize}

\section{Related Work}
Modeling human visual search process has been a long-standing topic in HCI and cognitive science. Several principles of human visual search mechanism have been identified, including the bottom-up processing and the top-down processing \cite{koch1987shifts}. The bottom-up processing refers to the finding that the salience of a visual stimulus is a major factor influencing attention deployment, and the salience is determined by the features of the stimuli (e.g., its color, shape, size and location) as well as the similarity between the stimuli and its surroundings. On the other hand, the top-down processing refers to the finding that attention deployment depends on the search task. In this case, the participants usually have task-relevant knowledge about the targets. For instance, the visual features of the target (e.g., its color, orientation and size) will guide the search so that the stimuli with matching features will receive more attention \cite{wolfe2017five}. Sometimes the explicit knowledge about the location of the target will also direct the attention \cite{jokinen2017modelling}.

Several computational models have utilized the idea of bottom-up and top-down processing to simulate eye-movement in visual search tasks \cite{kowler2011eye, tatler2005visual}. For example, Itti and Koch's model \cite{itti2000saliency} proposes that visual attention is allocated to the most salient visual region through a winner-take-all mechanism, i.e., the region that stands out the most from its surroundings. The model also incorporates a visual short-term memory component to prevent sustained attention to the most salient region. Particularly, the model implements an inhibitory mechanism to reduce the saliency of the recently attended region, i.e., inhibition of return. The idea of visual saliency, winner-take-all attention allocation, and inhibition of return are widely used in later computational models of visual search, e.g., \cite{jokinen2020adaptive}.

PAAV \cite{nyamsuren2013pre} is another computational model that implements bottom-up and top-down processing in visual search. It is based on the two-stage theory of visual processing: pre-attentive and attentive \cite{neisser1967cognitive, hoffman1979two}, and modeled after the computational \textit{guided search model} \cite{wolfe1994guided}. The pre-attentive process extracts information about the stimuli in the visual world in an automatic and parallel manner, including their locations, sizes, colors and orientations. Note that these features are detectable only when they exceeds their corresponding visual acuity thresholds. In addition to the  pre-attentive stage (bottom-up processing), the model has a top-down attentive stage, which is used to guide attention towards elements that share similar visual features with the desired target. In summary, the model presents a mathematical framework that can be used to compute bottom-up and top-down activations of a stimuli from its features that are visible at the current fixation, and attention is directed to the visual element with the largest weighted sum of these two kinds of activations.

Early research on visual search was often conducted in a restricted laboratory setting where the data was usually collected from well-controlled experiments with oversimplified visual stimuli. These experiments usually involved only a few independent variables, and thus the conclusions drawn from these experiments were also specific to the independent variables manipulated. Later studies have explored more realistic stimuli, such as menus \cite{bailly2014model}, grids \cite{Pfeuffer:2018:AMG:3173574.3173862} and webpages \cite{fu2007snif, teo2012cogtool}. Recently, Jokinen et al. \cite{jokinen2020adaptive} proposed a computational model of visual search on graphical layouts, with the assumption that the visual system is maximizing expected utility when deciding where to look at next. The work employed both the bottom-up processing and the top-down processing in the model when estimating the utility of different visual regions. Todi et al. \cite{todi2019individualising} uses a predictive model of visual search to optimize graphical layouts for an individual user so that items on an unseen interface can be found quicker. 

Despite being highly interpretable and even cognitively plausible, several problems remain unsolved in previous approaches. The first one is feature extraction. Previous models usually consider only a limited set of visual features such as location, color, shape, size and orientation. Furthermore, most of the time those features need to be extracted by the modeler beforehand and provided to the model symbolically \cite{jokinen2020adaptive}. In other words, the models do not deal with the problem of extracting visual features from raw pixels of a scene. Therefore, they cannot go beyond the limited set of primitive visual features, which are known in the literature or model designers, and cannot take advantage of the large amount of information that can be difficult to define or articulate. Although these handcrafted features are informative in search time prediction \cite{cockburn2007predictive, chen2015emergence}, unstructured features, which are ignored in previous work, might contain additional information in predicting search time and contribute uniquely to the complicated interaction between the target and the candidate stimuli in the visual scene. This is particularly relevant when the search tasks involve realistic stimuli like arbitrary webpages on the Internet. Second, many previous studies make strong assumptions about the task and the human behavior, e.g., links are organized in groups and human search hierarchically \cite{teo2012cogtool}. In reality, links are not always organized in groups and certain links may be so salient that they pop out immediately while others do not. Finally, previous models on realistic stimuli such as webpages are usually developed for particular target or visual scene, such as link search or menu selection. Our work aims to model a broad range of visual search tasks.

We employed a data-driven machine learning approach in our work, which does not make these assumptions about the task and the human behavior. One unique advantage of using deep learning is that it allows automatic extraction of useful features from unstructured data, which avoids labor-intensive feature engineering. In fact, several deep learning methods for modeling saliency map have been proposed (for a review, see \cite{borji2019saliency}). However, they are not intended for capturing goal-driven search tasks, i.e., users look for a specific target. A few recent papers have investigated goal-driven search tasks. For instance, \cite{Li:2018:PHP:3173574.3173603} presents a deep neural net to model human performance in performing a sequence of target selection tasks on a vertical list or menu. However, it remains a challenge to go beyond a specific type of UI such as menus and generalize a model to more challenging stimuli such as arbitrary webpages.  

Lastly, Zheng et al. \cite{zheng2018task} proposed a convolutional neural network to predict where people look at the webpage under different task conditions. Despite the promising results, several problems remain unsolved. First, they pre-defined five tasks, i.e., signing-up (email), information browsing (product promotion), form filling (file sharing, job searching), shopping (shopping) and community joining (social networking), and they pre-trained their model on a synthetic task-driven saliency dataset. Therefore, their model has been equipped with strong priors that are suitable to solve these pre-defined tasks, and on the other hand make it difficult to generalize to other tasks or to situations where priors are not available. In addition, the model only predicts the attention map, and not the search time. In contrast, our model does not assume the nature of the search task or the prior of the search pattern. It only requires the information about the webpage and the target, and directly predicts the search time. 

\begin{figure*}
  \centering
  \includegraphics[width=1.75\columnwidth]{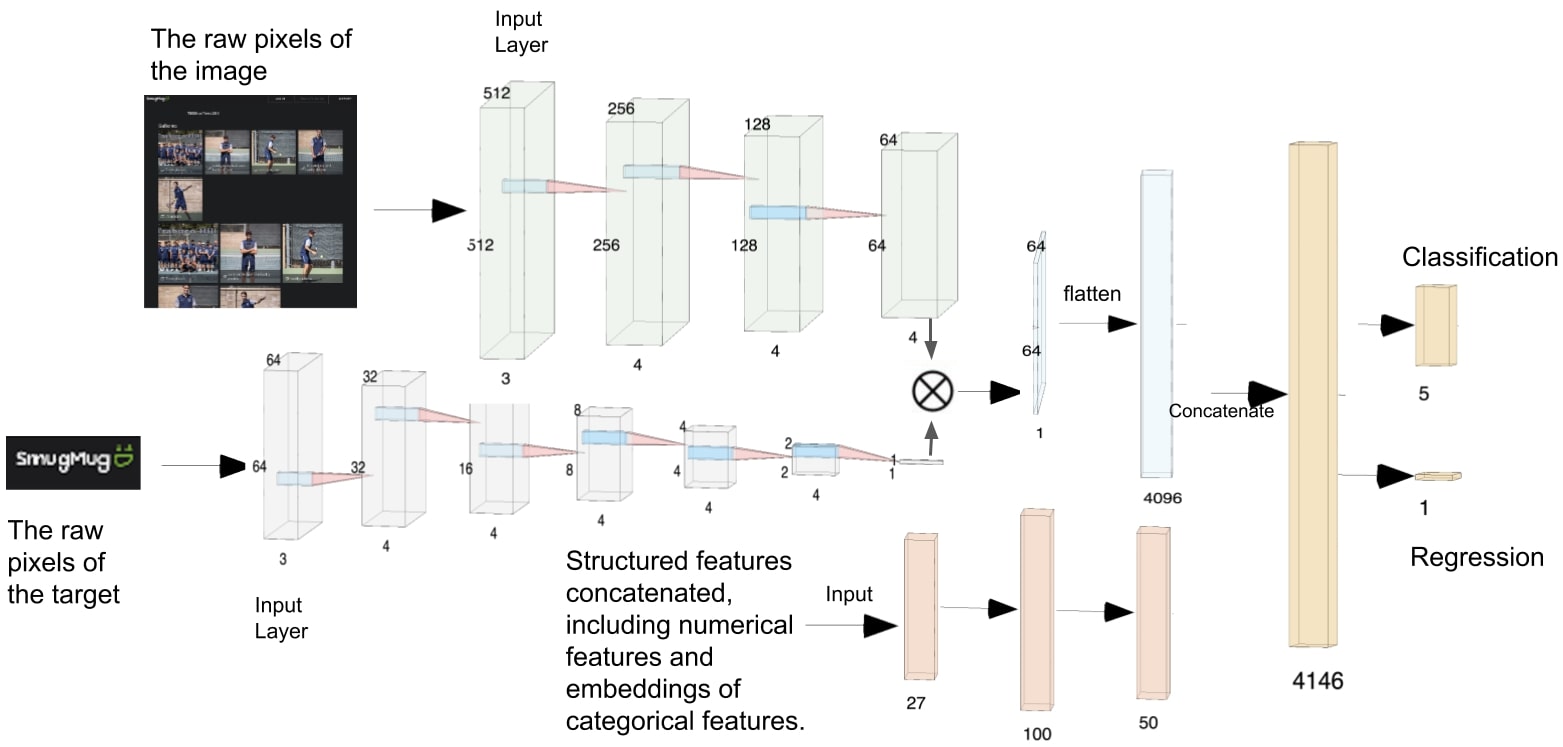}
  \caption{The model architecture. The attention map is computed as the alignment between the latent representations of the entire UI (the webpage) and the target, which is then concatenated with the structured features as the input for predicting human performance in visual search.}~\label{fig:model_architecture}
\end{figure*}

\section{THE MODEL DESIGN \& LEARNING}
We present a deep learning model for predicting human visual search performance on a webpage. In such a task, users are prompted to search a target on a realistic webpage, by locating and clicking on the target as quickly and accurately as possible. Although the task also involves components such as motor control, e.g., moving the mouse cursor and performing the actual click, we focus on the visual search aspect in this work. Previous work, e.g.,~\cite{bailly2014model}, has shown that visual search is a dominant component in the performance time. To build the model, we draw inspiration from cognitive psychology and cognitive neuroscience, and attempt to simulate the visual search process. In neuroscience, previous researchers have shown that when subjects try to selectively attend to a target in a visual scene, the neural representations of the objects that share similarity with the target (either the real target or the distractors) will all be enhanced \cite{corbetta2002control, martinez2004feature, liu2007feature}. We would like to capture the human attention mechanism with the neural attention algorithm in deep learning. Particularly, we utilizes the attention mechanism that has recently been widely adopted in image captaining and visual question answering \cite{anderson2018bottom}. Attention mechanism allows the model to selectively attend certain areas given the target. 
 In addition, since our model takes the pixel-level inputs, it is no longer necessary to hand-engineer specific features for different target types, e.g., image, text or button, since they now have a uniform representation of pixels. Finally, the model with neural attention mechanism allows us to examine the attention map it learns from the data and thus enhances the interpretability and the trustworthiness of the model.

The full architecture of our model is illustrated in Figure~\ref{fig:model_architecture}. Because targets can be of different sizes in different tasks, we first resize a target into the same dimension of $64 \times 64$. We then use a $6$-layer convolutional neural network to encode the target, which results in a target embedding of dimensions $1 \times 1 \times 4$. Similarly, for each task webpage, we first resize them to $512\times 512$ and then use a $3$-layer convolutional neural network to encode the webpage image from which we acquire an image embedding of $64 \times 64 \times 4$. Therefore, there are in total $64 \times 64$ super-pixel representations, each of which has dimension $4$. 

\subsection{Computing Alignment Scores}
With the embeddings of both the target (the goal) and the webpage images computed, we next perform a multiplicative attention mechanism \cite{xu2016ask, shih2016look} over these embeddings. In particular, we compute the cosine similarity between the target embedding vector and each super-pixel representation to get a 2-D attention map. We experiment with three design choices of the attention map: 1) using the original attention map; 2) applying the attention map to the image embedding to get attention-modulated image representation; and 3) normalizing the attention map with a $\mbox{softmax}(\cdot)$ function. In all these cases we treat the attention map as the representation of unstructured webpage-target features---referred as the webpage-target embedding---and combine it with heuristics-based, structured features (see next section), to predict search time. We find that the first method yields the best performance. Thus, we report the results based on the first formulation of attention map in the paper. The mathematical formula for computing the attention map $A \in \mathbb{R}^{64 \times 64}$ is the following:
\begin{equation}
A_{i,j} = \sum_{k=1}^{4} I_{i,j,k} T_k
\end{equation},
where $I \in \mathbb{R}^{64 \times 64 \times 4}$ is the output from the last convolutional layer of the webpage-CNN and $T \in \mathbb{R}^{4}$ is the output from the last convolutional layer of the target-CNN.

\subsection{Combining with Heuristic-Based Features}

To incorporate our intuition about factors that can affect visual search into the modeling process, we add several heuristic-based, structured features to the model. However, we ensure that these features are easily accessible and do not required domain-expertise in performance modeling. The features include: 1) the positional features, i.e., the (x, y) coordinates of the top left corner of the target bounding box; the Euclidean distance between the top left corner of the target and the top left corner of the screen; 2) the size features, i.e., the width, the height as well as the area of the target; 3) the number of candidates or distractors, i.e., the number of leaf nodes in the DOM representation of the webpage; and finally 4) target type. There are five possible target types: \texttt{image}, \texttt{text}, \texttt{link}, \texttt{button} and \texttt{input\_field}. 

Except for the target types, all the other features are treated as numerical variables. For target types, we regard it as a categorical variable. We use real-valued embedding vectors to encode different target types and the embeddings are learned from the data. The reason that we use embeddings instead of just one-hot representation of target types is that the five target types are not entirely mutually exclusive. For instance, ``link'' is a special type of text and most buttons also contain text. Using real-valued embedding vectors allows the model to fully capture the similarity and the differences between various target types, which we will discuss further later. 

After we combine all these structured features, including the original value of the numerical variables and the embedding of the categorical variable, we then feed them to a 2-layer fully connected network to get the feature embedding for the structured input. This embedding is then concatenated with the webpage-target embedding to jointly predict the search time using a final fully-connected layer. 

\subsection{Model Learning \& Loss Function}
Our model is implemented in Tensorflow \cite{abadi2016tensorflow}. It is trained through back-propagation, a typical method for training deep neural networks (see more details in the following sections). We examined different prediction tasks in our experiments as we will discuss later, and we train the model by minimizing a different loss function for each task: mean squared error loss for predicting continuous search time and cross entropy loss for predicting task difficulty categories. We use the Adam optimizer \cite{kingma2014adam} which have adaptive learning rates for model parameters. 
To avoid overfitting, we also add a L2-loss regularizer, which penalizes the model for having large connection weights. This is particularly useful when the training dataset is small, an issue that is often encountered in modeling human interaction performance. Another regularizer we use is the mean-squared-error between the attention map computed by the model and the binary mask generated by the target bounding box (e.g., Figure~\ref{fig:learned_attention}, right column). This provides intermediate supervision for the model and helps regularize the model parameters.

\section{Experiments}
We here first discuss our dataset, including data collection details and a data analysis. We then discuss our model experimentation based on the dataset.

\subsection{Data Collection \& Processing}
We experiment with our model based on a dataset collected via crowdsourcing. Webpages were randomly selected across the top 1 billion webpages, across 24 vertical classification categories. We only collected English webpages. For each webpage, a JSON hierarchy file of the webpage DOM tree and a PNG screenshot of the rendered webpage (normalized to $1024 \times 1024$) were collected. These webpages were then used to collect a dataset of search time labels from English-speaking crowd workers recruited via Amazon Mechanical Turk. 

\begin{figure}%
    \centering
    \subfloat[Searching for an image.]{{\includegraphics[width=1.0\linewidth]{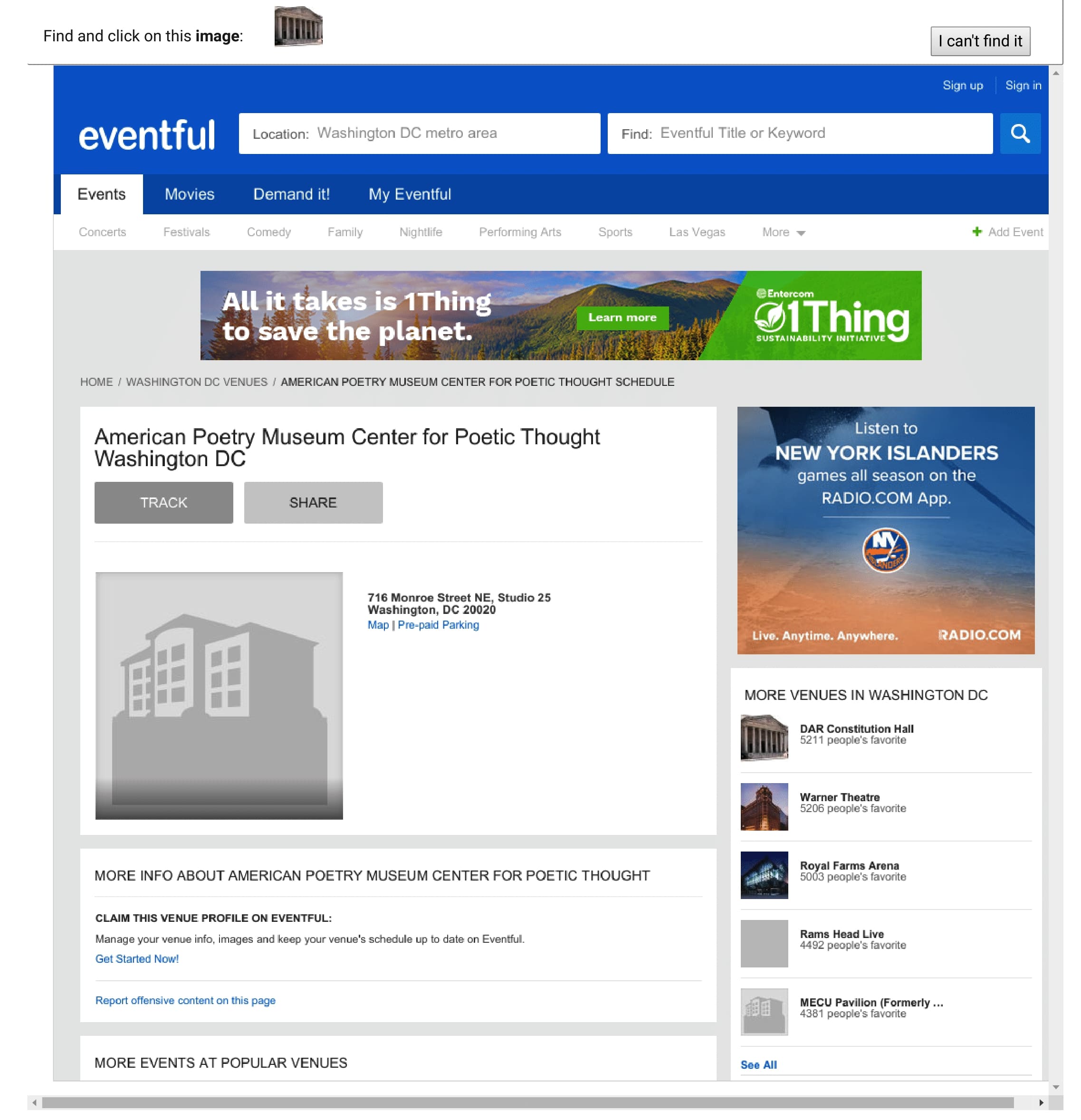} }}
    \qquad
    \subfloat[Searching for a link.]{{\includegraphics[width=1.0\linewidth]{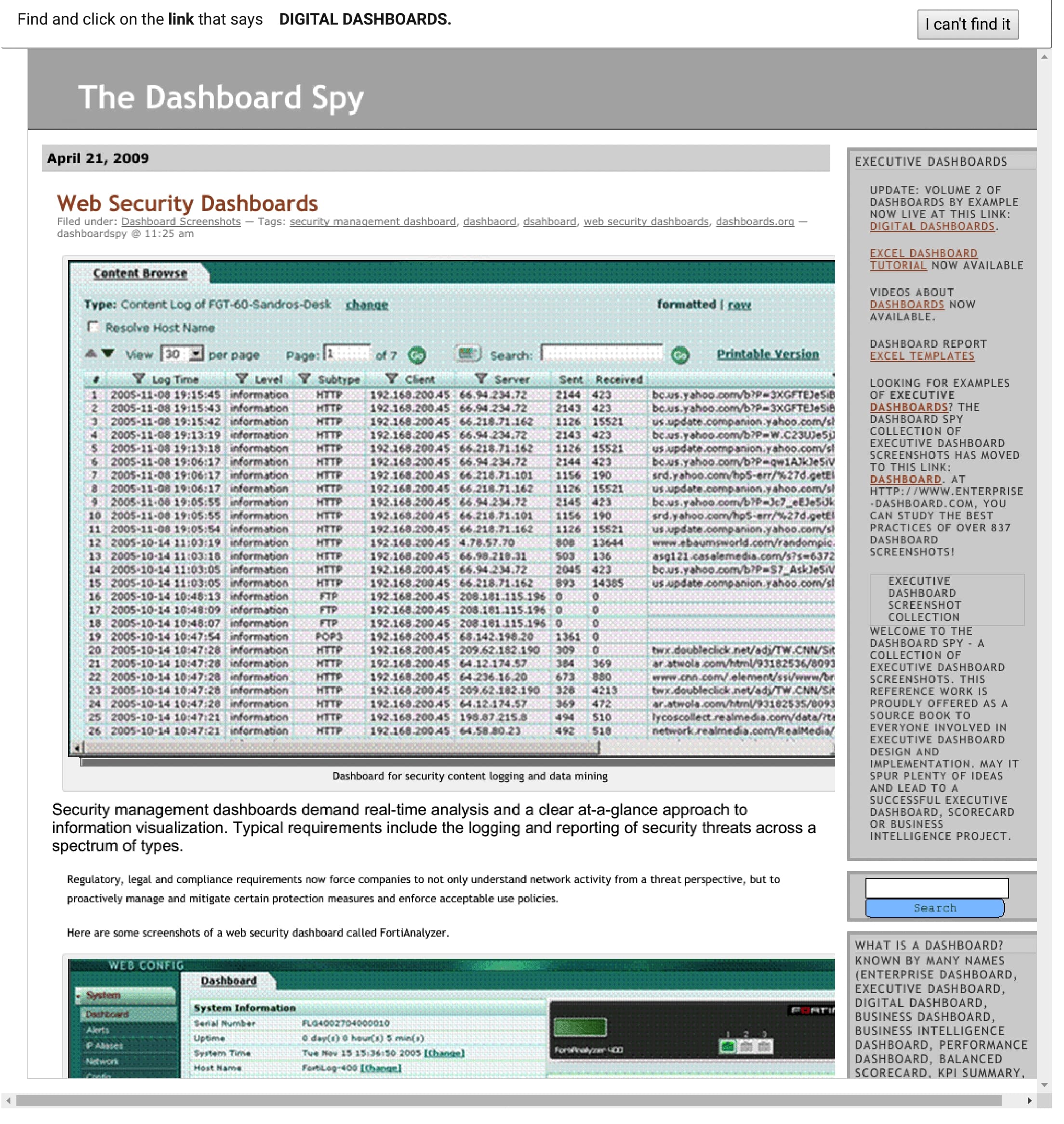} }}%
    \caption{Example screens in the crowdsourcing experiments. The target prompt was shown above the webpage, at the top of the screen.}%
    \label{fig:data_collection_sample}%
\end{figure}

For each task, an element on a webpage was randomly selected as the target, and the human worker was asked to find it on the webpage. We restrict a target element to be one of the following five types: \texttt{image}, \texttt{text}, \texttt{link}, \texttt{button} and \texttt{input\_field}. The minimum target width and height are 15px (an entire page is $1024 \times 1024$). We controlled the text complexity by limiting the maximum number of words in text targets to 5. 

Each trial starts by showing the human worker a screen that contains only the target prompt and a ``Begin'' button at the top left corner of the screen. Once the worker clicks on the ``Begin'' button, a webpage is revealed and the target prompt remains on the screen in case the worker forgets the target to look for. The trial is finished once the worker finds and clicks on the target, or clicks on the ``I can't find it'' button on the top right corner of the screen to skip the trial. Example tasks can be found in Figure~\ref{fig:data_collection_sample}.  If the target does not contain any text, its thumbnail image is shown to the user as the prompt for the search target (see Figure~\ref{fig:data_collection_sample}-a). Otherwise, the text content is displayed to the human worker for finding an element that contains such text (see Figure~\ref{fig:data_collection_sample}-b). In either case, the target is ensured to be a unique element on the page.

We excluded trials that are longer than 10 seconds as they are in the long tail of the time distribution (12\% of the entire dataset) and often outliers. For example, workers might be distracted for other events in their environment because these tasks were not conducted in a controlled environment. We asked workers to skip tasks where they have seen the webpages before (5.6\%) to remove learning effect, which can introduce a great variance in the search time as shown previously, e.g., \cite{bailly2014model}. We also filtered out the incorrect trials (10\%) in which users failed to click on the right target. We observed that incorrect trials tend to be longer (MEAN=8.7 seconds) than correct ones (MEAN=4.8 seconds), $t(39137) = 15.5, p=.000$. This indicates that incorrect trials cannot be explained by the speed-accuracy trade-off. Instead, we found those trials were indeed difficult and workers could resort to random guess after a long time of searching, e.g., targets in incorrect trials are significantly smaller than successful trials, t(39137)=-8.41, p=.000, which partially explains why participants failed the task. 

After these filtering steps, the total number of trials is 28,581, which were performed by 1,887 human workers. In the dataset, each user has at least 8 data points. We then randomly split the dataset for training ($1520$ users, $22591$ trials), validation ($184$ users, $3151$ trials) and testing ($183$ users, $2839$ trials). We ensure that there is no overlap in users among any of these datasets---the data of each user can only appear in one of these splits. 

\subsection{Analytic Examination of the Data}

To better understand the dataset and draw analytical understanding about human behaviors, we perform a set of analyses for the data regarding target positions, sizes and types as well as the complexity of the overall webpage. To aid our analysis, we normalize the search time of each data point by subtracting the mean of the entire dataset from it and then dividing it by the standard deviation. 

We use normalized search time because in our experiments each individual participant performed distinct search tasks with no repetition, which is a more challenging setup than previous setups. Previous work typically asked multiple users to perform the same task so that averaging time across participants would remove the individual difference and reduce noise for modeling. The normalization helps alleviate individual differences in our analyses, and it does not hurt the generalizability of the results since the ultimate answer we seek here is how difficult a task is relative to alternatives.

\subsubsection{Target Positions \& Sizes}
\begin{figure}[!ht]
\centering
  \includegraphics[width=0.95\columnwidth]{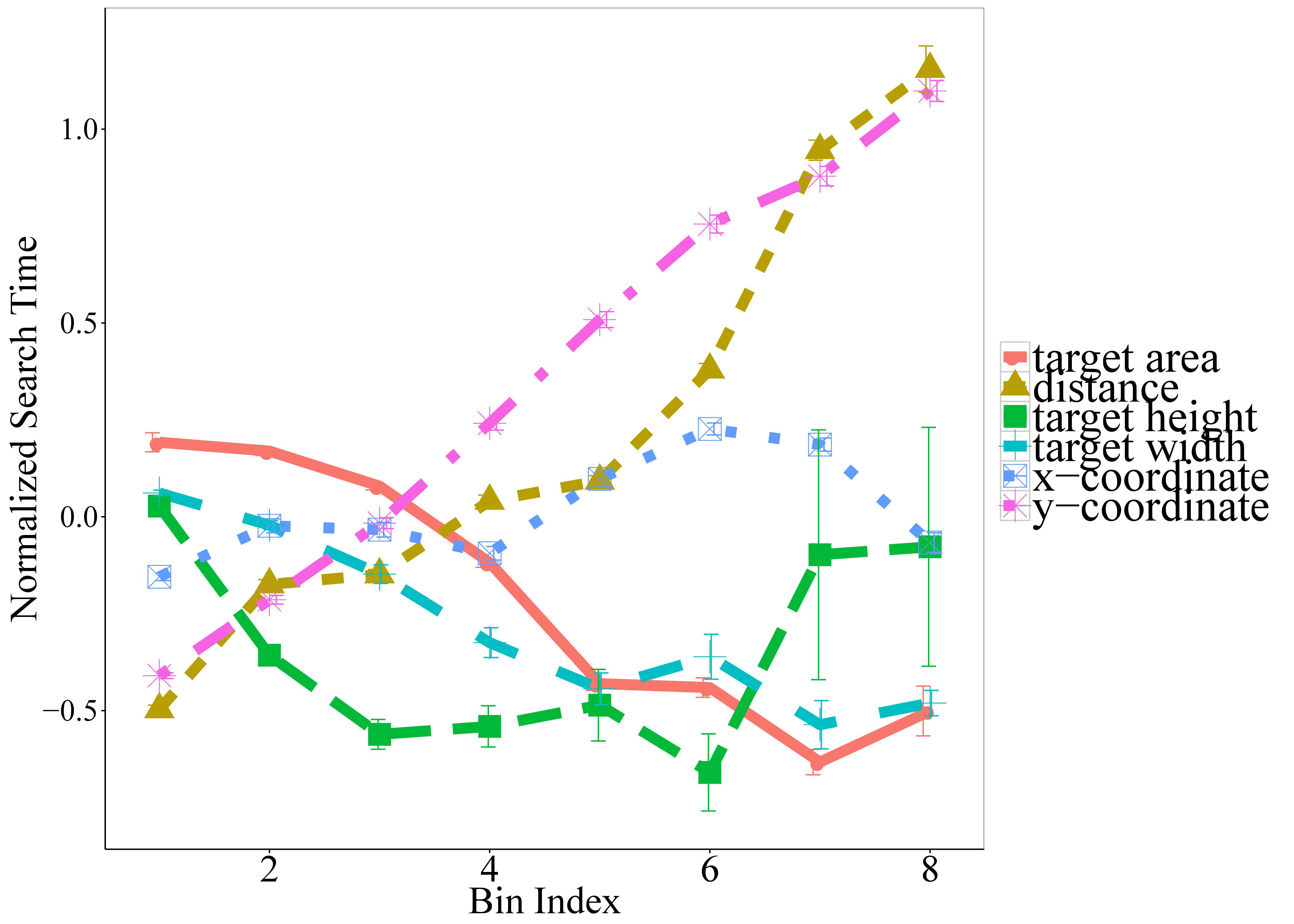}
  \caption{Search time over different positional features. Bin intervals for each feature are calculated by subtracting the minimum value from the maximum value of that feature and dividing by the number of bins. The error bars represent one standard error. }~\label{fig:position_factors}
\end{figure}

We first examine how the position of the target on a webpage affects search time. We find that the y-coordinate of the top left corner of the target is positively correlated with the search time: the bigger the y-coordinate, the longer it takes the participants to find the target, which implies a dominant top-down vertical search strategy employed by the human workers (see Figure~\ref{fig:position_factors}), since the starting position in our data is at the top of the screen. Regarding the size features, not surprisingly, we find that both the width and the height of the target negatively correlate with the search time, i.e., the bigger the target is, the shorter the search time. Based on a linear model that uses target vertical positions (y-coordinate) as one of the predictors, we found there is a positive correlation with the search time (see Table~\ref{tab:stats_numeirc}). In contrast, we found target area sizes negatively correlate with the search time. 
\begin{table}[ht]
\centering
\resizebox{\columnwidth}{!}{
\begin{tabular}{rrrrr}
  \hline
 & Coefficient & Std. Error & $t$ value & Pr($>$$|$t$|$) \\ 
  \hline
  y-coordinate & 0.2009 & 0.0035 & 57.79 & 0.0000 \\ 
  target area & -0.1105 & 0.0041 & -26.82 & 0.0000 \\ 
  total candidates & 0.1316 & 0.0073 & 18.06 & 0.0000 \\ 
   \hline
\end{tabular}
}
\caption{The statistical test of the linear model that uses target vertical position, target size and total number of objects to predict the normalized search time.} 
\label{tab:stats_numeirc}
\end{table}

\subsubsection{Page Complexities}
\begin{figure}[!ht]
\centering
  \includegraphics[width=0.95\columnwidth]{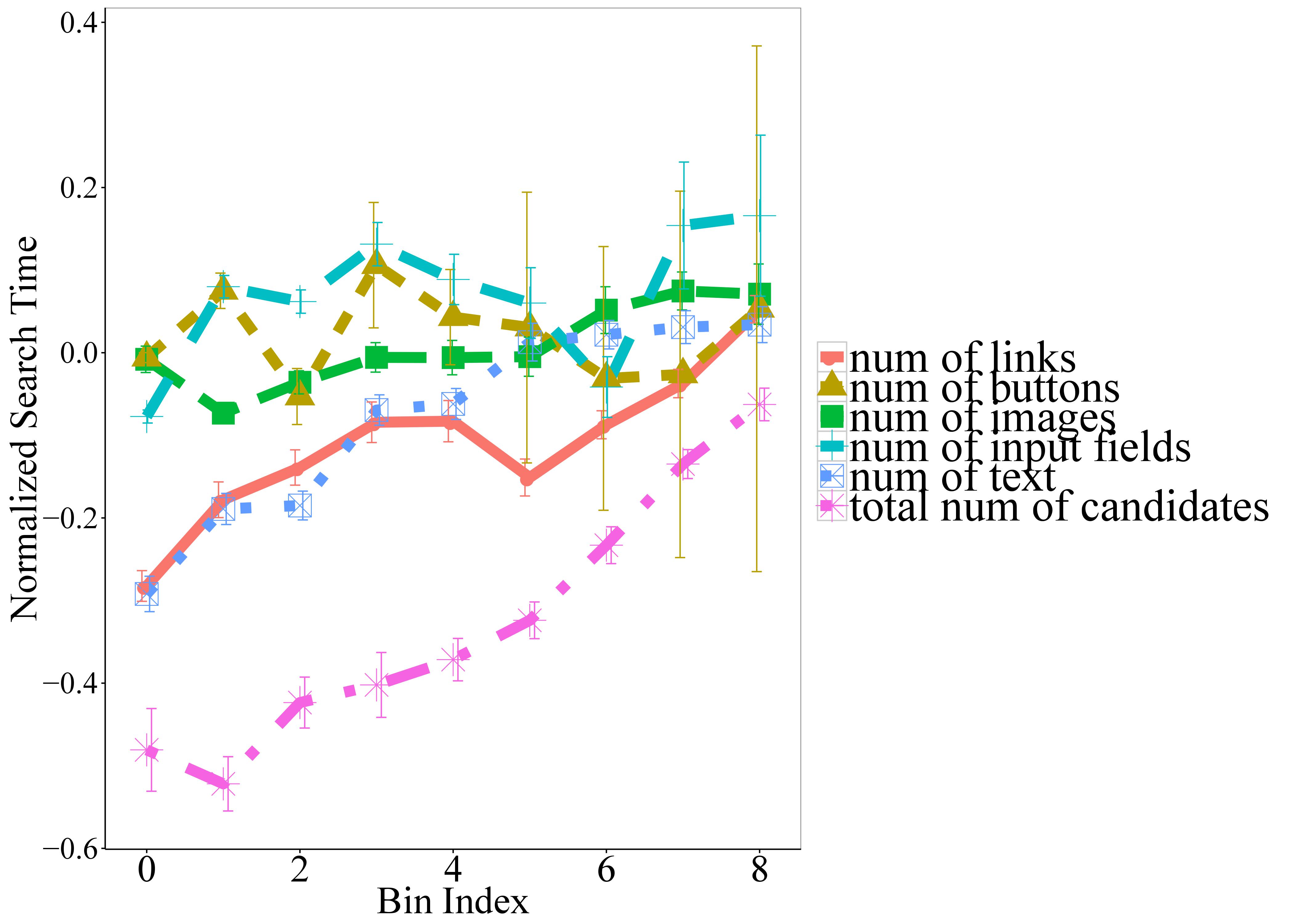}
  \caption{Search time over the total number of candidate targets on the page, and over the number of candidate targets of each type. Bin intervals for each feature are calculated by subtracting the minimum  value from the maximum value of that feature and dividing by the number of bins. The error bars represent one standard error.}~\label{fig:num_candidates_factors}
\end{figure}

We analyze the influence of the number of candidate items over the search time. Intuitively the more items on a page, the longer it should take a user to find a target. In addition to looking at how the total number of items affects search time, we conduct a finer granularity analysis by investigating how the number of candidate items in each object category would affect search time (see Figure~\ref{fig:num_candidates_factors}). We find that the number of certain object types are strong predictors of the search time. 

Taking into account Target Position, Size and Page Complexity, we run a linear model that use these numeric features to predict the search time. The $F$ statistics of this linear model is $F(3, 28577)=1910, p=.000$ and the coefficient estimates and the corresponding $p$ values for each feature can be found in Table~\ref{tab:stats_numeirc}.

\subsubsection{Target Types}
\begin{figure}
\centering
  \includegraphics[width=0.8\columnwidth]{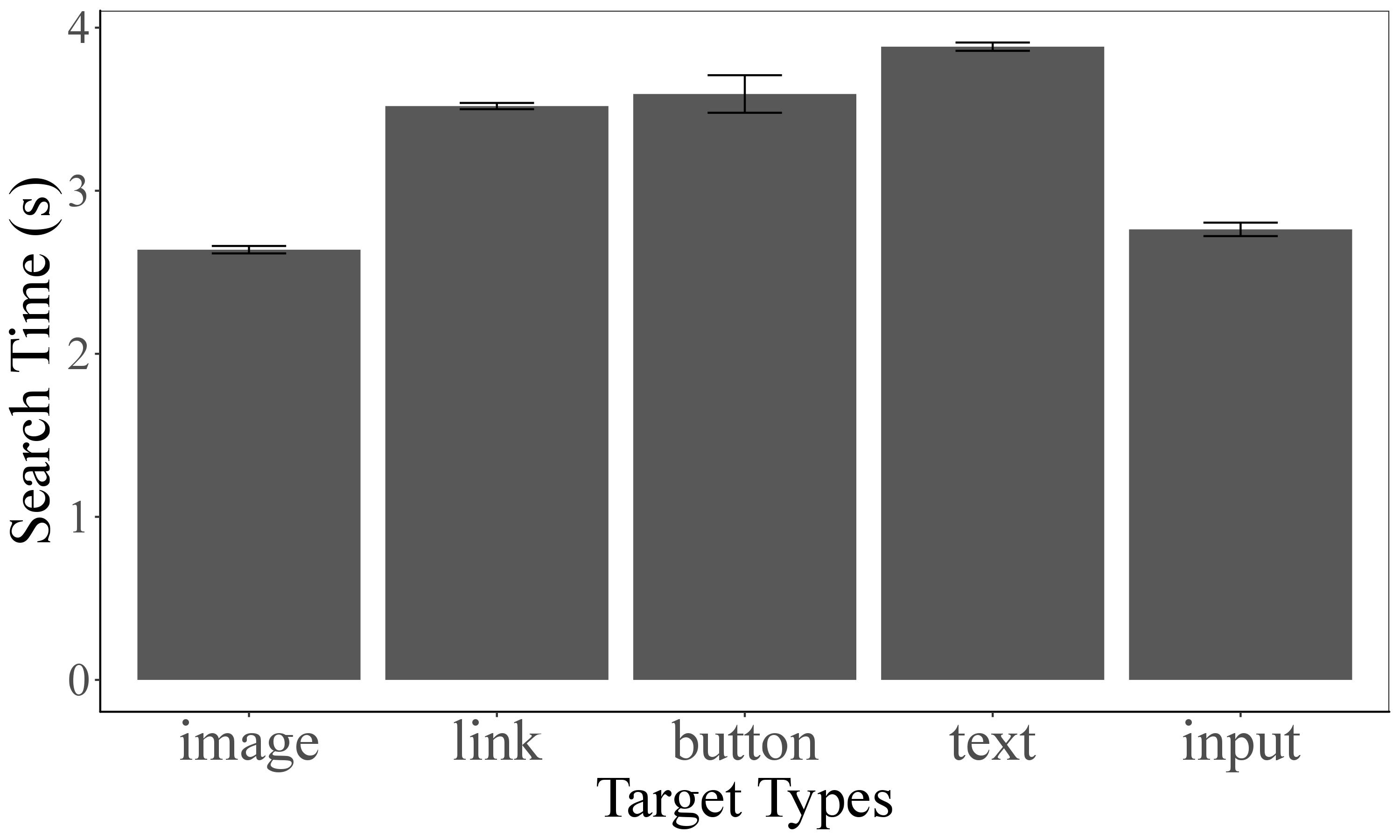}
  \caption{Search time over target types. Error bars represent one standard error. Images are the easiest to find whereas text is the hardest to find. See the main text for statistical analyses.}~\label{fig:target_types}
\end{figure}

As a categorical variable, Target Type has a significant impact on search time (Figure~\ref{fig:target_types}). For instance, \texttt{image} targets are the easiest to find and \texttt{text} targets are the hardest. Interestingly, \texttt{input\_field} targets are also easier to find, which was likely because users had more prior experience finding an input field than finding an arbitrary text on a webpage. The statistical test of the effect of target types are shown in Table~\ref{tab:stats_categorical}. The difference reflects the effect of each target type compared to the Image type. The $F$-statistics of the model is $F(4, 28576)=457, p=.000$. 

\begin{table}[ht]
\centering
\begin{tabular}{rrrrr}
  \hline
 & Difference & Std. Error & $t$ value & Pr($>$$|$t$|$) \\ 
  \hline
  link & 0.4500 & 0.0145 & 31.06 & 0.0000 \\ 
  button & 0.5164 & 0.0518 & 9.97 & 0.0000 \\ 
  text & 0.6222 & 0.0159 & 39.18 & 0.0000 \\ 
  input & 0.0767 & 0.0247 & 3.11 & 0.0019 \\ 
   \hline
\end{tabular}
\caption{The statistical test of the linear model that compares the normalized search time of different target types. The Difference is computed by subtracting the mean normalized search time of the Image type from that of each other type.} 
\label{tab:stats_categorical}
\end{table}

\begin{table}[ht]
\centering
\resizebox{\columnwidth}{!}{%
\begin{tabular}{lllll}
  \hline
Model & within-user $R^2$ & cross-user $R^2$ & classification & ranking \\ 
  \hline
x-coordinate & 0.113 & 0.011 & 0.254 & 0.542 \\ 
  y-coordinate & 0.338 & 0.229 & 0.346 & 0.669 \\ 
  target width & 0.097 & 0.015 & 0.235 & 0.524 \\ 
  target height & 0.124 & 0.025 & 0.279 & 0.601 \\ 
  distance & 0.177 & 0.089 & 0.294 & 0.622 \\ 
  target area & 0.106 & 0.037 & 0.266 & 0.567 \\ 
  target type & 0.103 & 0.012 & 0.250 & 0.551 \\ 
  total candidates & 0.137 & 0.051 & 0.276 & 0.578 \\ 
  structured-all & 0.373 & 0.267 & 0.355 & 0.691 \\ 
  \textbf{deep net+structured} & \textbf{0.384**} & \textbf{0.288***} & \textbf{0.366***} & \textbf{0.699***} \\ 
   \hline
\end{tabular}
}
\caption{The performances of different models. The first eight rows (all but the last two rows) report the results of linear models using a single feature (indicated by the Model column). The second last row (structured-all) refers to the baseline linear model that uses all the structured features. The last row reports the performance of our model, which uses both structured and unstructured inputs.} 
\label{tab:results}
\end{table}

\subsection{Model Configuration \& Hyperparameters}

The webpage-CNN has $3$ convolutional layers (see Figure~\ref{fig:model_architecture}) where each layer has a kernel size of $3 \times 3$ and output channels of $4$. Between the convolutional layers we also add a batch normalization layer \cite{ioffe2015batch} to stabilize and accelerate training, a ReLU activation function to introduce non-linearity to the model, and a max-pooling layer with pool size $2 \times 2$. The target-CNN has the same architecture excepts that it uses 6 convolutional layers. The webpage embedding computed by the webpage-CNN is $64 \times 64 \times 4$ and the target embedding computed by the target-CNN is $1 \times 1 \times 4$. We then compute the cosine similarity between the target embedding and each of the super-pixel representation in the webpage embedding. The saliency map is thus $64 \times 64$. The salience map is then flattened to yield an embedding of dimension $4096$.

On the other hand, the structured features are obtained from concatenating the continuous variables and the embedding of target type, a categorical variable. The dimension of the embedding is $20$, so the total dimension of the combined structured features is $27$. This feature vector is fed into a fully-connected neural network with two hidden layers. The numbers of hidden units are $100$ and $50$ and the activation function is tanh. The output of the fully connected layer, a $50$-dimension vector, is then concatenated with the webpage-target saliency embedding ($4096$-dim). To predict a continuous time value, the concatenation is fed to a fully-connected layer with a single output unit, with no activation function. 

We also evaluate our model on two additional tasks: classification and ranking tasks. For classification task, we categorize a search task into five classes: Very-Easy, Easy, Neutral, Hard, and Very Hard. We bucketize a continuous search time into one of these five categories according to the percentiles within participants. Specifically, for each participant, we label the data points smaller than the 20th percentile of that user's data as Very-Easy, the data points between the 20th and the 40th percentile as Easy, between the 40th and the 60th percentile as Neutral, between the 60th and the 80th percentile as Hard and above the 80th percentile as Very-Hard. In this way, we convert the original regression problem into a classification problem and we also guarantee that each label class has roughly same number of data points, so that the dataset is balanced. To solve the classification task, we replace the single output unit with a 5-unit softmax output layer, representing the probability of a search task belonging to one of these categories. For the ranking metric, we simply use the prediction from the regression model to rank the difficulty of two search tasks.

We train the model by minimizing the loss using the Adam optimizer, an algorithm for first-order gradient-based optimization of stochastic objective functions \cite{kingma2014adam}, with a learning rate of
0.0001. The batch size is $8$. Because we do not want the model to produce an attention map that only highlights the target, we assign a small weight (0.001) to the mean-squared-error loss capturing the difference between the attention map and the binary mask generated by the target bounding box. To regularize the model
learning, we also apply a dropout ratio of $10$\% to all the
fully-connected layers, which prevents the model from overfitting \cite{srivastava2014dropout}.

\subsection{Performance Results}
We use the validation dataset to determine the optimal stopping time and the engineering decisions we make during the model iteration. Previous modeling work in HCI or cognitive psychology often reports performance on how well a model can fit the data. As a deep learning neural network has a vast number of parameters, it is trivial to fit the training data perfectly. Therefore, in this paper, we only report the model results on the test dataset, which the model has no access to during the training time and the authors have no access to during the model development. This truly reflects how the modeling architecture is capable of modeling unseen data, which is also more realistic in everyday application. 

For the regression task, we report the $R^2$ metric for both within and cross-user cases. For the within-user case, we compute the $R^2$ over the trials of each user and then average the scores over all the users.  For the cross-user case, we directly compute $R^2$ over all the trials from all the users. Both of them measure the correlation between the predicted and the ground-truth visual search times. We also report the classification accuracy and the ranking accuracy (Table~\ref{tab:results}). As we discussed earlier, there was no repeat in performing each task both within and across users, and all these metrics are computed based on the performance of each unique task, instead of the average performance of multiple trials---the approach often used in traditional modeling work. Thus, it is extremely challenging to obtain high accuracy due to noises and individual differences. 

From Table~\ref{tab:results}, We can see that among all the single-feature baseline models, y-coordinate has the highest performance. The baseline model that uses all the structured features (structured-all) performs the best among all the baseline models. Our deep model (the last row in the table) that combines both structured and unstructured input outperforms all the baseline models that use only structured features across all the three metrics. Particularly, the within-user $R^2$ of our model is significantly greater than the structured-all baseline model, $t(4) = 4, p<.01$. The cross-user  $R^2$ of our model is also significantly greater, $t(4) = 8, p<.001$. The 5-way classification accuracy is significantly greater than the structured-all model, $t(4) = 9, p<.001$. The same holds true for the ranking accuracy, $t(4) = 8, p<.001$.

\begin{figure*}[t]
    \centering
    \subfloat[Searching for an image.]{{\includegraphics[width=0.9\linewidth]{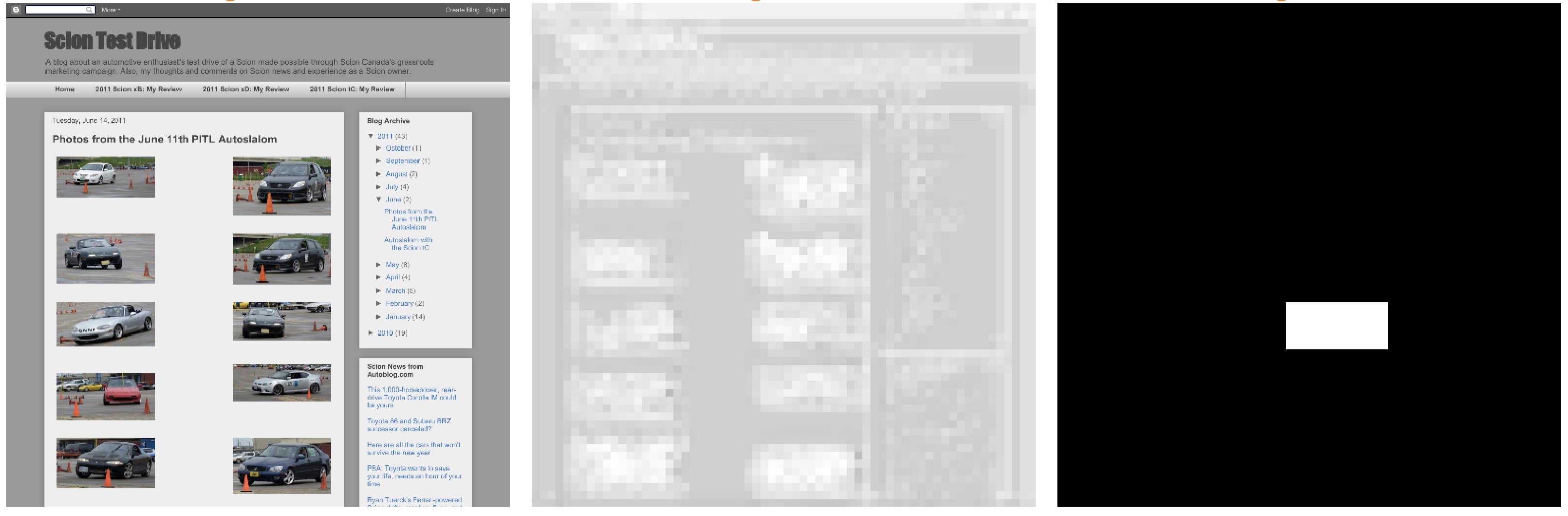} }}%
    \qquad
    \subfloat[Searching for a link.]{{\includegraphics[width=0.9\linewidth]{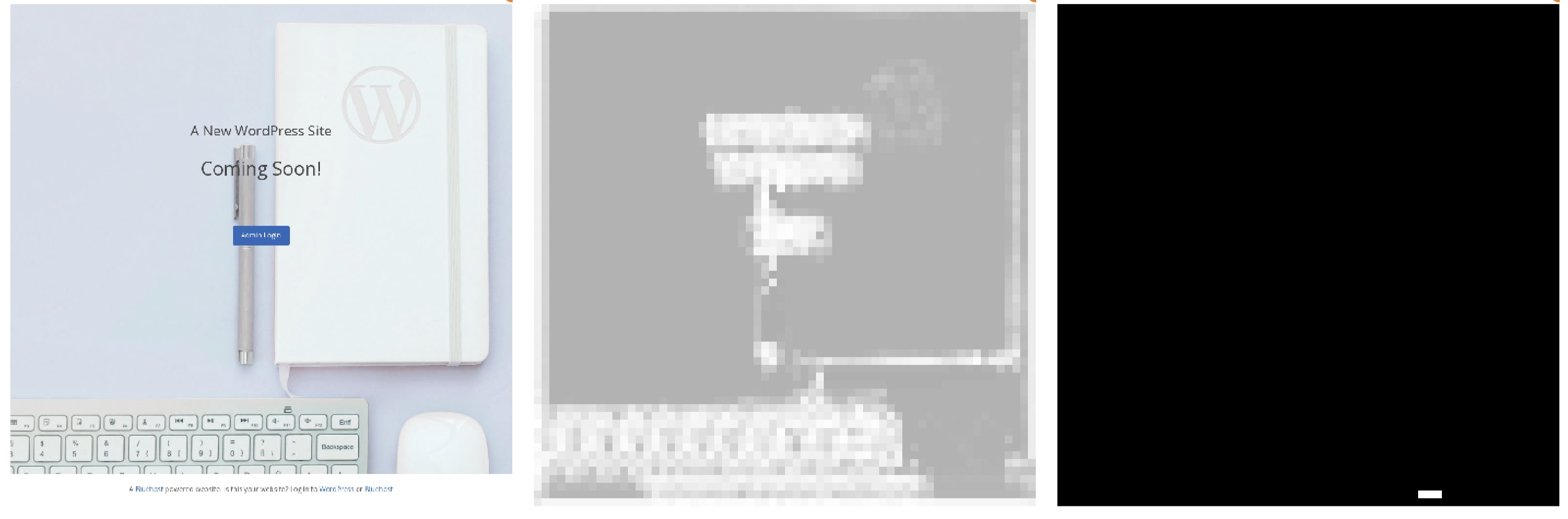} }}%
    \caption{The attention map visualization of two example tasks. The left figures show the webpage of each task. The middle figures show the corresponding goal-driven attention map. The right figures show the ground truth location of the target in each task.}%
    \label{fig:learned_attention}%
\end{figure*}

\begin{figure*}%
    \centering
    \subfloat[Classification Task.]{{\includegraphics[width=0.45\linewidth]{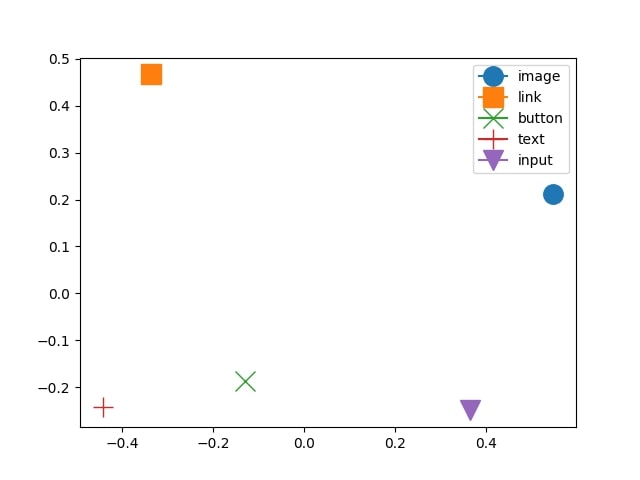} }}%
    \subfloat[Regression Task]{{\includegraphics[width=0.45\linewidth]{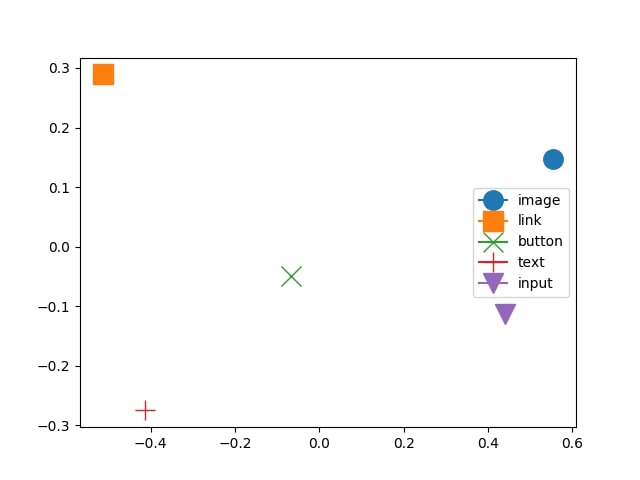} }}%
    \caption{The PCA projection of embeddings of each target type.}%
    \label{fig:embedding_analysis}%
\end{figure*}

\section{Analyzing Model Behaviors}
To better understand the model prediction, we examine the attention map learned by the model, a latent representation computed via Equation 1. We found that the model is able to detect content on the webpage, regardless of the background color of the webpage. As we can see in Figure~\ref{fig:learned_attention}, target bounding box usually falls into one of the highlighted regions in the attention map. Interestingly, the attention map tends to capture the distractors that are visually similar to the target. For instance, in Figure~\ref{fig:learned_attention}-a, the target is an image and the model learns to highlight the images and ignore the text on the page, whereas in Figure~\ref{fig:learned_attention}-b, when the target is text, the model learns to highlight text but suppress images in the attention map.

To gain a deeper understanding of the model behavior, we analyzed the embeddings for each of the 5 target types learned by the model. The embedding dimension for each target type is $20$, and we project these dimensions onto a 2-dimensional space using Principle Component Analysis (PCA) for visualization  (see Figure~\ref{fig:embedding_analysis}). We found sensible structures in these embeddings. For the first principle component, i.e., the x-axis in Figure~\ref{fig:embedding_analysis}, \texttt{link} and \texttt{text} types are closer than others because both types are text-based. They are followed by \texttt{button}, which often contain text but not always so. They are then followed by \texttt{input\_field} and finally \texttt{image}. In other words, there is a sensible transition from text-like stimuli to image-like stimuli. This kind of property emerged both in regression task training and in classification task training. 

Note that these embedding relationships are automatically formed as part of the learning. In our training, the model was not directly guided to form these embedding relationships. Instead, these patterns naturally occur as the model learns from the data because targets of specific types tend to manifest similar difficulties for search tasks thus are closer in the embedding space.

\section{Discussion \& Future Work} 
Convolutional neural networks with attentional mechanism have been widely used in computer vision tasks, such as visual question answering or image captioning \cite{yang2016stacked, lu2016hierarchical}. The attention mechanism is particularly useful when there is a need to selectively attend to a small region in the image. The reason that we use attentional CNN to model visual search time is that it naturally computes the similarity between the target and the candidates on the webpage, thus giving us a saliency map that captures both the presence of the target and the distractors. This design is also biologically motivated \cite{liu2007feature, treue1999feature, martinez2004feature}. However, if we only use attentional CNN to process the raw pixels of the webpages (unstructured data), it is not enough to achieve a good performance, because we are omitting the structured features that are known to predict search time but difficult to distill through raw pixels without using an even larger dataset. For examples, the influence of high-level concepts such as target types could be where users brought their prior experience to the task, which are difficult to learn only through raw pixels. In contrast, they can be easily encoded as a feature. Therefore, we extend our deep learning framework to flexibly integrate both structured and unstructured features and learn the weights for them from the data. 

The benefit of deep learning approaches for human performance modeling is multi-fold. First, through learning from the data, the model naturally captures the priors that humans have, such as the priority of the top left quadrant of the webpage compared to other regions. Second, when it comes to complicated tasks, such as searching a target on a busy webpage, the interaction between the target and the background cannot be simply characterized by a single or a few handcrafted parameters. Third, traditional linear models do not support modeling complicated interactions between features. In contrast, the great capacity of deep learning models allow us to easily capture complex relationships among structured and unstructured features.  

There are multiple applications of the current work. First, if designers want to add a new element to a webpage and optimize its search time, they could move the element around to compare the predicted search time for different locations. In another situation, the model can be used to restructure graphical layouts so that the important items on the webpage have the shortest averaged search time. Last but not least, the model can be used for saliency editing, making an item more salient on a webpage. Particularly, the designer can experiment with different design choices, e.g., the font size, color, and shape of a button, in order to design a target such that it leads to the shortest search time. Although it cannot automate a design, the model can aid the designer in an iterative design process.

There are several limitations with the current work. First, since the ``Begin'' button is always positioned at the top left corner of the screen, the starting position of the search is the same across trials, which is different from real life scenarios. We believe our current model is still useful as it can be used to examine the relative performance of alternative designs via A/B testing when the starting position is controlled. Because our approach is data driven, this limitation can be easily addressed by collecting trials with different starting positions in future research. Another limitation is model interpretability, a common problem with most deep learning approaches. In this paper, we examine the attention map learned by the model to understand what the model has learned. We find that in many cases they are able to selectively attend to potential targets, which is similar to human selective attention. We also analyze the embeddings of target types learned by the model, and find that it naturally captures the properties of different target types, e.g., a spectrum from text-like stimuli to image-like stimuli spontaneously emerges. These findings help us better understand model predictions and make the model more explainable. In future work, we intend to further analyze model behaviors with various model architectures to better understand how the model learns to perform the task.

Despite being biologically inspired, our model is not intended to replicate the real human visual search process. For one thing, it does not involve any sequential attention shifts, which is different from human behavior. In future work we will utilize deep learning models such as recurrent neural networks to capture sequential aspects of visual search. This would allow us to model human search behaviors at a finer granularity by which we can compare model prediction and human data on a more detailed level. Nevertheless, our investigation in this work offers valuable insights into both how well and how a deep model would approach the task.

There are several other directions for future work. For example, although our model outperforms the baseline models with structured features, there is still a lot of room for improvement in terms of model performance. In addition, we currently only focus on search time. In future work, we can extend the model to give designers suggestions on how to reduce search time by improving target saliency under design constraints. 

\section{Conclusion}

While visual search has been extensively studied, our work adds to the body of literature by modeling visual search without making any assumptions on the visual environment (arbitrary webpages). It provides a unified framework to model diverse search tasks, which can flexibly integrate both structured and unstructured features. Our results show that combining both structured and unstructured features gives the model unique advantage to both fit the training data and generalize to the unseen data. It outperforms baseline models that use only structured features. Our methodology can be readily extended to incorporate other cognitive models and heuristics as well as emerging deep learning models to combine the strength of each method. To understand what the model has learned, we examine the latent representations learned by the model and observed spontaneous emergence of sensible representations of target type as well as intuitive attention map. Our modeling work demonstrates the potential of using deep learning neural networks to model realistic visual search in human-computer interaction. 

\section{Acknowledgments}
We would like to thank Amanda Swearngin for her help on the early data collection, and Xin Zhou for his feedback on the infrastructure implementation.

\balance{}

% BALANCE COLUMNS
\balance{}

% REFERENCES FORMAT
% References must be the same font size as other body text.
\bibliographystyle{SIGCHI-Reference-Format}
\bibliography{sample}


\begin{thebibliography}{00}

%%% the Web address.

\ifx \showCODEN    \undefined \def \showCODEN     #1{\unskip}     \fi
\ifx \showDOI      \undefined \def \showDOI       #1{{\tt DOI:}\penalty0{#1}\ }
  \fi
\ifx \showISBNx    \undefined \def \showISBNx     #1{\unskip}     \fi
\ifx \showISBNxiii \undefined \def \showISBNxiii  #1{\unskip}     \fi
\ifx \showISSN     \undefined \def \showISSN      #1{\unskip}     \fi
\ifx \showLCCN     \undefined \def \showLCCN      #1{\unskip}     \fi
\ifx \shownote     \undefined \def \shownote      #1{#1}          \fi
\ifx \showarticletitle \undefined \def \showarticletitle #1{#1}   \fi
\ifx \showURL      \undefined \def \showURL       #1{#1}          \fi

\bibitem{abadi2016tensorflow}
{Mart{\'\i}n Abadi}, {Paul Barham}, {Jianmin Chen}, {Zhifeng Chen}, {Andy
  Davis}, {Jeffrey Dean}, {Matthieu Devin}, {Sanjay Ghemawat}, {Geoffrey
  Irving}, {Michael Isard}, {Manjunath Kudlur}, {Josh Levenberg}, {Rajat
  Monga}, {Sherry Moore}, {Derek~G. Murray}, {Benoit Steiner}, {Paul Tucker},
  {Vijay Vasudevan}, {Pete Warden}, {Martin Wicke}, {Yuan Yu}, {and} {Xiaoqiang
  Zheng}. 2016.
\newblock \showarticletitle{TensorFlow: A System for Large-Scale Machine
  Learning}. In {\em 12th {USENIX} Symposium on Operating Systems Design and
  Implementation ({OSDI} 16)}. {USENIX} Association, Savannah, GA, 265--283.
\newblock
\showISBNx{978-1-931971-33-1}


\bibitem{anderson2018bottom}
{Peter Anderson}, {Xiaodong He}, {Chris Buehler}, {Damien Teney}, {Mark
  Johnson}, {Stephen Gould}, {and} {Lei Zhang}. 2018.
\newblock \showarticletitle{Bottom-up and top-down attention for image
  captioning and visual question answering}. In {\em Proceedings of the IEEE
  Conference on Computer Vision and Pattern Recognition}. 6077--6086.
\newblock


\bibitem{bailly2014model}
{Gilles Bailly}, {Antti Oulasvirta}, {Duncan~P Brumby}, {and} {Andrew Howes}.
  2014.
\newblock \showarticletitle{Model of visual search and selection time in linear
  menus}. In {\em Proceedings of the SIGCHI Conference on Human Factors in
  Computing Systems}. ACM, 3865--3874.
\newblock


\bibitem{borji2019saliency}
{Ali Borji}. 2019.
\newblock \showarticletitle{Saliency Prediction in the Deep Learning Era:
  Successes and Limitations}.
\newblock {\em IEEE Transactions on Pattern Analysis and Machine
  Intelligence\/} (2019).
\newblock


\bibitem{chen2015abc}
{Kan Chen}, {Jiang Wang}, {Liang-Chieh Chen}, {Haoyuan Gao}, {Wei Xu}, {and}
  {Ram Nevatia}. 2015b.
\newblock \showarticletitle{ABC-CNN: An Attention Based Convolutional Neural
  Network for Visual Question Answering}.
\newblock {\em arXiv preprint arXiv:1511.05960\/} (2015).
\newblock


\bibitem{chen2015emergence}
{X Chen}, {G Bailly}, {DP Brumby}, {A Oulasvirta}, {and} {A Howes}. 2015a.
\newblock \showarticletitle{The Emergence of Interactive Behaviour: A Model of
  Rational Menu Search}. In {\em CHI'15 Proceedings of the 33rd Annual ACM
  Conference on Human Factors in Computing Systems}, Vol.~33. Association for
  Computing Machinery (ACM), 4217--4226.
\newblock


\bibitem{cockburn2007predictive}
{Andy Cockburn}, {Carl Gutwin}, {and} {Saul Greenberg}. 2007.
\newblock \showarticletitle{A Predictive model of Menu Performance}. In {\em
  Proceedings of the SIGCHI conference on Human Factors in Computing Systems}.
  ACM, 627--636.
\newblock


\bibitem{corbetta2002control}
{Maurizio Corbetta} {and} {Gordon~L Shulman}. 2002.
\newblock \showarticletitle{Control of goal-directed and stimulus-driven
  attention in the brain}.
\newblock {\em Nature Reviews Neuroscience\/} {3}, 3 (2002), 201.
\newblock


\bibitem{devlin2018bert}
{Jacob Devlin}, {Ming-Wei Chang}, {Kenton Lee}, {and} {Kristina Toutanova}.
  2018.
\newblock \showarticletitle{BERT: Pre-training of deep bidirectional
  transformers for language understanding}.
\newblock {\em arXiv preprint arXiv:1810.04805\/} (2018).
\newblock


\bibitem{fu2007snif}
{Wai-Tat Fu} {and} {Peter Pirolli}. 2007.
\newblock \showarticletitle{SNIF-ACT: A cognitive model of user navigation on
  the World Wide Web}.
\newblock {\em Human--Computer Interaction\/} {22}, 4 (2007), 355--412.
\newblock


\bibitem{hoffman1979two}
{James~E Hoffman}. 1979.
\newblock \showarticletitle{A two-stage model of visual search}.
\newblock {\em Perception \& Psychophysics\/} {25}, 4 (1979), 319--327.
\newblock


\bibitem{ioffe2015batch}
{Sergey Ioffe} {and} {Christian Szegedy}. 2015.
\newblock \showarticletitle{Batch normalization: Accelerating deep network
  training by reducing internal covariate shift}.
\newblock {\em arXiv preprint arXiv:1502.03167\/} (2015).
\newblock


\bibitem{itti2000saliency}
{Laurent Itti} {and} {Christof Koch}. 2000.
\newblock \showarticletitle{A saliency-based search mechanism for overt and
  covert shifts of visual attention}.
\newblock {\em Vision Research\/} {40}, 10-12 (2000), 1489--1506.
\newblock


\bibitem{johnson2017google}
{Melvin Johnson}, {Mike Schuster}, {Quoc~V Le}, {Maxim Krikun}, {Yonghui Wu},
  {Zhifeng Chen}, {Nikhil Thorat}, {Fernanda Vi{\'e}gas}, {Martin Wattenberg},
  {Greg Corrado}, {and} {others}. 2017.
\newblock \showarticletitle{Google’s multilingual neural machine translation
  system: Enabling zero-shot translation}.
\newblock {\em Transactions of the Association for Computational Linguistics\/}
   {5} (2017), 339--351.
\newblock


\bibitem{jokinen2017modelling}
{Jussi~PP Jokinen}, {Sayan Sarcar}, {Antti Oulasvirta}, {Chaklam
  Silpasuwanchai}, {Zhenxin Wang}, {and} {Xiangshi Ren}. 2017.
\newblock \showarticletitle{Modelling learning of new keyboard layouts}. In
  {\em Proceedings of the 2017 CHI Conference on Human Factors in Computing
  Systems}. ACM, 4203--4215.
\newblock


\bibitem{jokinen2020adaptive}
{Jussi~PP Jokinen}, {Zhenxin Wang}, {Sayan Sarcar}, {Antti Oulasvirta}, {and}
  {Xiangshi Ren}. 2020.
\newblock \showarticletitle{Adaptive feature guidance: Modelling visual search
  with graphical layouts}.
\newblock {\em International Journal of Human-Computer Studies\/}  {136}
  (2020), 102376.
\newblock


\bibitem{kingma2014adam}
{Diederik~P Kingma} {and} {Jimmy Ba}. 2014.
\newblock \showarticletitle{Adam: A method for stochastic optimization}.
\newblock {\em arXiv preprint arXiv:1412.6980\/} (2014).
\newblock


\bibitem{koch1987shifts}
{Christof Koch} {and} {Shimon Ullman}. 1987.
\newblock \showarticletitle{Shifts in selective visual attention: towards the
  underlying neural circuitry}.
\newblock In {\em Matters of Intelligence}. Springer, 115--141.
\newblock


\bibitem{kowler2011eye}
{Eileen Kowler}. 2011.
\newblock \showarticletitle{Eye movements: The past 25 years}.
\newblock {\em Vision Research\/} {51}, 13 (2011), 1457--1483.
\newblock


\bibitem{krizhevsky2012imagenet}
{Alex Krizhevsky}, {Ilya Sutskever}, {and} {Geoffrey~E Hinton}. 2012.
\newblock \showarticletitle{ImageNet classification with deep convolutional
  neural networks}. In {\em Advances in Neural Information Processing Systems}.
  1097--1105.
\newblock


\bibitem{lecun2015deep}
{Yann LeCun}, {Yoshua Bengio}, {and} {Geoffrey Hinton}. 2015.
\newblock \showarticletitle{Deep Learning}.
\newblock {\em Nature\/} {521}, 7553 (2015), 436.
\newblock


\bibitem{Li:2018:PHP:3173574.3173603}
{Yang Li}, {Samy Bengio}, {and} {Gilles Bailly}. 2018.
\newblock \showarticletitle{Predicting Human Performance in Vertical Menu
  Selection Using Deep Learning}. In {\em Proceedings of the 2018 CHI
  Conference on Human Factors in Computing Systems}. ACM, New York, NY, USA,
  Article 29, 7 pages.
\newblock
\showISBNx{978-1-4503-5620-6}


\bibitem{liu2007feature}
{Taosheng Liu}, {Jonas Larsson}, {and} {Marisa Carrasco}. 2007.
\newblock \showarticletitle{Feature-based attention modulates
  orientation-selective responses in human visual cortex}.
\newblock {\em Neuron\/} {55}, 2 (2007), 313--323.
\newblock


\bibitem{lu2016hierarchical}
{Jiasen Lu}, {Jianwei Yang}, {Dhruv Batra}, {and} {Devi Parikh}. 2016.
\newblock \showarticletitle{Hierarchical question-image co-attention for visual
  question answering}. In {\em Advances In Neural Information Processing
  Systems}. 289--297.
\newblock


\bibitem{martinez2004feature}
{Julio~C Martinez-Trujillo} {and} {Stefan Treue}. 2004.
\newblock \showarticletitle{Feature-based attention increases the selectivity
  of population responses in primate visual cortex}.
\newblock {\em Current Biology\/} {14}, 9 (2004), 744--751.
\newblock


\bibitem{mcelree1999temporal}
{Brian McElree} {and} {Marisa Carrasco}. 1999.
\newblock \showarticletitle{The temporal dynamics of visual search: evidence
  for parallel processing in feature and conjunction searches.}
\newblock {\em Journal of Experimental Psychology: Human Perception and
  Performance\/} {25}, 6 (1999), 1517.
\newblock


\bibitem{neisser1967cognitive}
{Ubric Neisser}. 1967.
\newblock \showarticletitle{Cognitive Psychology (New York: Appleton)}.
\newblock {\em Century, Crofts\/} (1967).
\newblock


\bibitem{nyamsuren2013pre}
{Enkhbold Nyamsuren} {and} {Niels~A Taatgen}. 2013.
\newblock \showarticletitle{Pre-attentive and attentive vision module}.
\newblock {\em Cognitive Systems Research\/}  {24} (2013), 62--71.
\newblock


\bibitem{Pfeuffer:2018:AMG:3173574.3173862}
{Ken Pfeuffer} {and} {Yang Li}. 2018.
\newblock \showarticletitle{Analysis and Modeling of Grid Performance on
  Touchscreen Mobile Devices}. In {\em Proceedings of the 2018 CHI Conference
  on Human Factors in Computing Systems}. ACM, New York, NY, USA, Article 288,
  12 pages.
\newblock
\showISBNx{978-1-4503-5620-6}


\bibitem{ren2015faster}
{Shaoqing Ren}, {Kaiming He}, {Ross Girshick}, {and} {Jian Sun}. 2015.
\newblock \showarticletitle{Faster R-CNN: Towards real-time object detection
  with region proposal networks}. In {\em Advances in Neural Information
  Processing Systems}. 91--99.
\newblock


\bibitem{shen2003guidance}
{Jiye Shen}, {Eyal~M Reingold}, {and} {Marc Pomplun}. 2003.
\newblock \showarticletitle{Guidance of eye movements during conjunctive visual
  search: the distractor-ratio effect.}
\newblock {\em Canadian Journal of Experimental Psychology\/} {57}, 2 (2003),
  76.
\newblock


\bibitem{shih2016look}
{Kevin~J Shih}, {Saurabh Singh}, {and} {Derek Hoiem}. 2016.
\newblock \showarticletitle{Where to look: Focus regions for visual question
  answering}. In {\em Proceedings of the IEEE Conference on Computer Vision and
  Pattern Recognition}. 4613--4621.
\newblock


\bibitem{srivastava2014dropout}
{Nitish Srivastava}, {Geoffrey Hinton}, {Alex Krizhevsky}, {Ilya Sutskever},
  {and} {Ruslan Salakhutdinov}. 2014.
\newblock \showarticletitle{Dropout: a simple way to prevent neural networks
  from overfitting}.
\newblock {\em The Journal of Machine Learning Research\/} {15}, 1 (2014),
  1929--1958.
\newblock


\bibitem{tatler2005visual}
{Benjamin~W Tatler}, {Roland~J Baddeley}, {and} {Iain~D Gilchrist}. 2005.
\newblock \showarticletitle{Visual correlates of fixation selection: Effects of
  scale and time}.
\newblock {\em Vision Research\/} {45}, 5 (2005), 643--659.
\newblock


\bibitem{tehranchi2018modeling}
{Farnaz Tehranchi} {and} {Frank~E Ritter}. 2018.
\newblock \showarticletitle{Modeling visual search in interactive graphic
  interfaces: Adding visual pattern matching algorithms to ACT-R}. In {\em
  Proceedings of 16th International Conference on Cognitive Modeling}.
  University of Wisconsin Madison, WI, 162--167.
\newblock


\bibitem{teo2012cogtool}
{Leong-Hwee Teo}, {Bonnie John}, {and} {Marilyn Blackmon}. 2012.
\newblock \showarticletitle{CogTool-Explorer: a model of goal-directed user
  exploration that considers information layout}. In {\em Proceedings of the
  SIGCHI Conference on Human Factors in Computing Systems}. ACM, 2479--2488.
\newblock


\bibitem{todi2019individualising}
{Kashyap Todi}, {Jussi Jokinen}, {Kris Luyten}, {and} {Antti Oulasvirta}. 2019.
\newblock \showarticletitle{Individualising Graphical Layouts with Predictive
  Visual Search Models}.
\newblock {\em ACM Transactions on Interactive Intelligent Systems (TiiS)\/}
  {10}, 1 (2019), 1--24.
\newblock


\bibitem{treisman1980feature}
{Anne~M Treisman} {and} {Garry Gelade}. 1980.
\newblock \showarticletitle{A feature-integration theory of attention}.
\newblock {\em Cognitive Psychology\/} {12}, 1 (1980), 97--136.
\newblock


\bibitem{treue1999feature}
{Stefan Treue} {and} {Julio C~Martinez Trujillo}. 1999.
\newblock \showarticletitle{Feature-based attention influences motion
  processing gain in macaque visual cortex}.
\newblock {\em Nature\/} {399}, 6736 (1999), 575.
\newblock


\bibitem{van2016towards}
{Hidde van~der Meulen}, {Petra Varsanyi}, {Lauren Westendorf}, {Andrew~L Kun},
  {and} {Orit Shaer}. 2016.
\newblock \showarticletitle{Towards understanding collaboration around
  interactive surfaces: Exploring joint visual attention}. In {\em Proceedings
  of the 29th Annual Symposium on User Interface Software and Technology}. ACM,
  219--220.
\newblock


\bibitem{Walter:2015:AVA:2750858.2804255}
{Robert Walter}, {Andreas Bulling}, {David Lindlbauer}, {Martin Schuessler},
  {and} {J\"{o}rg M\"{u}ller}. 2015.
\newblock \showarticletitle{Analyzing Visual Attention During Whole Body
  Interaction with Public Displays}. In {\em Proceedings of the 2015 ACM
  International Joint Conference on Pervasive and Ubiquitous Computing}. ACM,
  New York, NY, USA, 1263--1267.
\newblock
\showISBNx{978-1-4503-3574-4}


\bibitem{wolfe1994guided}
{Jeremy~M Wolfe}. 1994.
\newblock \showarticletitle{Guided search 2.0 a revised model of visual
  search}.
\newblock {\em Psychonomic Bulletin \& Review\/} {1}, 2 (1994), 202--238.
\newblock


\bibitem{wolfe2017five}
{Jeremy~M Wolfe} {and} {Todd~S Horowitz}. 2017.
\newblock \showarticletitle{Five factors that guide attention in visual
  search}.
\newblock {\em Nature Human Behaviour\/} {1}, 3 (2017), 0058.
\newblock


\bibitem{wu2018analysis}
{Xiaoli Wu}, {Tom Gedeon}, {and} {Linlin Wang}. 2018.
\newblock \showarticletitle{The analysis method of visual information searching
  in the human-computer interactive process of intelligent control system}. In
  {\em Congress of the International Ergonomics Association}. Springer, 73--84.
\newblock


\bibitem{xu2016ask}
{Huijuan Xu} {and} {Kate Saenko}. 2016.
\newblock \showarticletitle{Ask, attend and answer: Exploring question-guided
  spatial attention for visual question answering}. In {\em European Conference
  on Computer Vision}. Springer, 451--466.
\newblock


\bibitem{yang2016stacked}
{Zichao Yang}, {Xiaodong He}, {Jianfeng Gao}, {Li Deng}, {and} {Alex Smola}.
  2016.
\newblock \showarticletitle{Stacked attention networks for image question
  answering}. In {\em Proceedings of the IEEE Conference on Computer Vision and
  Pattern Recognition}. 21--29.
\newblock


\bibitem{zhaoping2011clash}
{Li Zhaoping} {and} {Uta Frith}. 2011.
\newblock \showarticletitle{A clash of bottom-up and top-down processes in
  visual search: The reversed letter effect revisited.}
\newblock {\em Journal of Experimental Psychology: Human Perception and
  Performance\/} {37}, 4 (2011), 997.
\newblock


\bibitem{zheng2018task}
{Quanlong Zheng}, {Jianbo Jiao}, {Ying Cao}, {and} {Rynson~WH Lau}. 2018.
\newblock \showarticletitle{Task-driven webpage saliency}. In {\em Proceedings
  of the European Conference on Computer Vision (ECCV)}. 287--302.
\newblock


\end{thebibliography}

\end{document}